\documentclass[12pt,preprint]{aastex}
\shorttitle{Virgo Globular Clusters}
\shortauthors{Strader et al.}
\def\etal{{\it et al.}}

\begin{document}

\title{Globular Clusters in Virgo Ellipticals: Unexpected Results for Giants, Dwarfs, and Nuclei from ACS Imaging}

\author{Jay Strader, Jean P. Brodie, Lee Spitler, Michael A. Beasley}
\affil{UCO/Lick Observatory, University of California, Santa Cruz, CA 95064}
\email{strader@ucolick.org, brodie@ucolick.org, lees@ucolick.org, mbeasley@ucolick.org}

\begin{abstract}

We have analyzed archival \emph{Hubble Space Telescope}/Advanced Camera for Surveys images in $g$ and $z$ of 
the globular cluster (GC) systems of 53 ellipticals in the Virgo Cluster, spanning massive galaxies to dwarf 
ellipticals (dEs). Several new results emerged: (i) In the giant ellipticals (gEs) M87 and NGC 4649, there is 
a correlation between luminosity and color for \emph{individual} metal-poor GCs, such that more massive GCs 
are more metal-rich. A plausible interpretation of this result is self-enrichment, and may suggest that these 
GCs once possessed dark matter halos. (ii) In some gEs (most notably M87), there is an ``interloping'' 
subpopulation of GCs with intermediate colors (1.0 $< g-z <$ 1.25) and a narrow magnitude range (0.5 mag) 
near the turnover of the GC luminosity function. These GCs look otherwise identical to the classic metal-poor 
and metal-rich GC subpopulations. (iii) The dispersion in color is nearly twice as large for the metal-rich 
GCs than the metal-poor GCs. However, there is evidence for a nonlinear relation between $g-z$ and 
metallicity, and the dispersion in metallicity may be the same for both subpopulations. (iv) Very luminous, 
intermediate-color GCs are common in gEs. These objects may be remnants of many stripped dwarfs, analogues of 
$\omega$ Cen in the Galaxy. (v) There is a continuity of GC system colors from gEs to some dEs: in 
particular, many dEs have metal-rich GC subpopulations. We also confirm the GC color--galaxy luminosity 
relations found previously for both metal-poor and metal-rich GC subpopulations. (vi) There are large 
differences in GC specific frequency among dEs, independent of the presence of a nucleus and the fraction of 
metal-rich GCs. Over $-15 < M_B < -18$, there is little correlation between specific frequency and $M_B$ (in 
contrast to previous studies). But we do find evidence for two separate $S_N$ classes of dEs: those with 
$B$-band $S_N \sim 2$, and dEs with populous GC systems that have $S_N$ ranging from $\sim 5-20$ with median 
$S_N \sim 10$. Together, these points suggest multiple formation channels for dEs in the Virgo Cluster. (vii) 
The peak of the GC luminosity function (GCLF) is the same for both gEs and dEs. This is contrary to 
expectations of dynamical friction on massive GCs, unless the primordial GCLF varies between gEs and dEs. 
Among gEs the GCLF turnover varies by a surprising small 0.05 mag, an encouraging result for its use as an 
accurate standard candle. (viii) dE,Ns appear bimodal in their nuclear properties: there are small bright red 
nuclei consistent with formation by dynamical friction of GCs, and larger faint blue nuclei which appear to 
have formed by a dissipative process with little contribution from GCs. The role of dynamical evolution in 
shaping the present-day properties of dE GC systems and their nuclei remains ambiguous.

\end{abstract}


\section{Introduction}

It is increasingly apparent that globular clusters (GCs) offer important constraints on the star 
formation and assembly histories of galaxies. Recent spectroscopic studies of GCs in massive 
early-type galaxies (e.g., Strader \etal~2005) indicate that the bulk of star formation occurred 
at relatively high redshift ($z \ga 2$) in high density environments (as environmental density and 
galaxy mass decrease, the fraction of younger GCs may increase; see Puzia \etal~2005). These 
findings allow the age-metallicity degeneracy to be broken and lead to the conclusion that the 
bimodal color distributions seen in most nearby luminous galaxies are due primarily to two old GC 
subpopulations: metal-poor (blue) and metal-rich (red). The metallicities of these peaks correlate 
with host galaxy luminosity (Larsen \etal~2001, Strader, Brodie, \& Forbes 2004; see also Lotz, 
Miller, \& Ferguson 2004 for dwarfs).

Most recent photometric studies of GC systems in ellipticals have used the Wide Field and 
Planetary Camera 2 (WFPC2) on the Hubble Space Telescope (HST). Compared to ground-based imaging, 
this strategy gains photometric accuracy and minimizes contamination at the expense of small 
spatial coverage. Among the larger HST studies of early-type galaxies utilizing deep imaging are 
Larsen \etal~(2001) and Kundu \& Whitmore (2001). These studies found bimodality in many of their 
sample galaxies (extending down to low-luminosity ellipticals) and a nearly uniform log-normal GC 
luminosity function (GCLF) with a peak at $M_V \sim -7.4$. However, the GC systems of dwarf 
ellipticals (dEs) are more poorly understood. The large HST surveys to date (primarily of Virgo 
and Fornax) have been limited to relatively shallow snapshot imaging; this precluded the study of 
color and luminosity distributions in detail. Among the suggestions of this initial work are a 
correlation of increasing specific frequency ($S_N$) with decreasing galaxy luminosity, a 
dichotomy in the GC systems of nucleated and non-nucleated dEs (dE,N and dE,noN, respectively), 
and the difficulty of making dE nuclei as observed through dynamical friction of GCs (Miller 
\etal~1998, Lotz \etal~2001).

Data from the Advanced Camera for Surveys (ACS) Virgo Cluster Survey (HST GO 9401, P.~I.~C{\^ 
o}t{\' e}) offers an important step forward in understanding the detailed properties of the GC 
systems of ellipticals over a wide range in galaxy mass. The use of F475W and F850LP filters 
(henceforth called $g$ and $z$ for convenience, though the filters do not precisely match the 
Sloan ones) allows a much wider spectral baseline for metallicity separation than $V$ and $I$. 
Though only a single orbit is used per galaxy, the increased sensitivity of ACS (compared to 
WFPC2) allows one to reach $\sim 3$ mag beyond the turnover of the GCLF, encompassing most of the 
GCs in a given galaxy. More accurate photometry for the brighter GCs is also possible. Finally, 
the field of view of ACS is twice that of WFPC2. Together, these attributes allow a study of the 
color and luminosity distribution of GCs in a large sample of galaxies in much more detail than 
previously possible.

In what follows, elliptical (unabbreviated) refers to all galaxies in our sample and denotes no 
specific luminosity. The three brightest galaxies are described as giant ellipticals (gEs); these 
have $M_B \le -21.4$. Galaxies of intermediate or high luminosity are called Es. Faint galaxies 
with exponential surface brightness profiles are dEs; the transition from E to dE occurs 
traditionally at $M_B \sim -18$ (Kormendy 1985). We include two galaxies with $M_B = -18.1$ (VCC 
1422 and VCC 1261) under this heading, since these galaxies have nuclei similar to those commonly 
found among dEs. The E/dE classifications have been taken from the literature and we do not 
perform independent surface photometry in this paper (though we note the increasing debate in the 
literature about whether this dichotomy is real, e.g., Graham \& Guzm{\' a}n 2003). We have 
updated the nucleation status of a dE if appropriate; the vast majority of the dEs in our study 
have nuclei.

\section{Data Reduction and Analysis}

All data were taken as part of the ACS Virgo Cluster Survey (C{\^ o}t{\' e} \etal~2004); this 
survey includes both ellipticals and S0s. We used all galaxies classified as ellipticals, 
excepting a few dwarfs quite close to luminous Es whose GC systems could not be isolated. This 
left a final sample of 53 galaxies. Images were first processed through the standard ACS pipeline. 
\emph{Multidrizzle} was utilized for image combining and cosmic ray rejection. GC candidates were 
selected as matched-filter detections on 20 $\times$ 20 pixel median-subtracted images. Using 
DAOPHOT II (Stetson 1993), aperture photometry was performed in a 5-pixel aperture and adjusted to 
a 10-pixel aperture using corrections of $-0.09$ in $g$ and $-0.15$ in $z$. These are median
corrections derived from bright objects in the five most luminous galaxies in the Virgo 
Cluster Survey: VCC 1226, VCC 1316, VCC 1978, VCC 881, and VCC 798. These 10-pixel magnitudes were 
then corrected to a nominal infinite aperture using values of $-0.10$ in $g$ and $-0.12$ in $z$ (Sirianni 
\etal~2005; this paper describes the photometric calibration of ACS). Finally, the magnitudes were 
transformed to the AB 
system using zeropoints from Sirianni \etal~(26.068 and 24.862 for $g$ and $z$, respectively), and 
corrected for Galactic reddening using the maps of Schlegel Finkbeiner, \& Davis (1998).

Most GCs at the distance of Virgo are well-resolved in ACS imaging. Half-light radii ($r_h$) for 
GC candidates were measured on $g$ images (since the $g$ PSF is more centrally concentrated) using 
the \emph{ishape} routine (Larsen 1999). For each object, King models with fixed $c=30$ (for 
$c=r_{tidal}/r_{core}$) and varying $r_h$ were convolved with a distant-dependent empirical PSF 
derived from bright isolated stars in the images to find the best-fit $r_h$. This $c$ is typical 
of non core-collapsed GCs in the Milky Way (Trager, King, \& Djorgovski 1995). We experimented 
with allowing $c$ to vary, but it was poorly constrained for most GCs. However, the adopted 
$c$ in \emph{ishape} has little effect on the derived $r_h$ (Larsen 1999). To convert these 
measured sizes into physical units, galaxy distance estimates are required. We used those derived 
from surface brightness fluctuation measurements in the literature when possible: these were 
available from Tonry \etal~(2001) for the bright galaxies and from Jerjen \etal~(2004) for several 
dEs. For the remainder of the galaxies we used a fixed distance of 17 Mpc, which is the mean of 
the ellipticals in Tonry \etal

Due to the depth of the images ($z \ga 25$), some of the fields suffer significant contamination 
from foreground stars and especially background galaxies. Using the gEs and several of the more
populous dEs as fiducials, we chose the following structural cuts to reduce interlopers:  0.55 $<$ sharp $<$ 0.9, 
$-0.5 <$ round $< 0.5$, and $1 < r_h$ (pc) $< 13$, where the sharp and round parameters are from 
DAOPHOT. A large upper limit for $r_h$ is used since the size measurements skew systematically larger for 
fainter GCs. We further applied a color cut of $0.5 < g-z < 2.0$ ($> 0.3$ mag to each blue and red of 
the limiting metallicities expected for old GCs;  Jord{\' a}n \etal~2004) and an error limit $<$ 
0.15 mag. In practice, this magnitude limit excluded most GCs within the innermost few arcsec of 
the brightest galaxies (whose GC systems are quite populous). Finally, we visually inspected all 
GC candidates, and excluded those which were obviously background galaxies. Our criteria are illustrated
visually in Figure 1 for the bright dE VCC 1087, which displays a good mix of actual GCs and contaminants.

These cuts remove nearly all foreground stars. However, compact galaxies (or compact star-forming 
regions within larger galaxies) with the appropriate colors can masquerade as GCs. In some images, 
clusters of galaxies are clearly visible. The increasing numbers of background sources at $z \ga 
23$, combined with the difficulty of accurate size measurements below this magnitude, makes 
efficient rejection of contaminants challenging. For gEs this is a minimal problem, due to the 
large number of GCs within the ACS field of view (hundreds to $\sim 1700$ for M87). But with dEs 
and even low-luminosity Es, contaminants can represent a large fraction of the GC candidates. Due 
to these concerns, we chose only to use GCs brighter than $z = 23.5$ to study the colors and total 
numbers of GCs for the remainder of the paper. However, we used minimal cuts to study GC luminosity functions;
this is described in more detail in \S 4.

Basic data about the 53 galaxies in our sample are given in Table 1, along with GC system 
information as discussed below.

\section{Color Distributions}

\subsection{Massive Ellipticals}

In Figures 2 and 3 we show the color-magnitude diagrams (CMDs) for the three most luminous 
galaxies in our sample: NGC 4472 ($M_B = -21.9$), M87 ($M_B = -21.5$), and NGC 4649 ($M_B = 
-21.4$). Figure 4 is a plot of magnitude vs.~photometric error in $g$ and $z$ for M87. The CMDs in 
Figures 2 and 3 contain considerable structure only apparent because of the large number of GCs; 
we have chosen to discuss them in some detail.

All three gEs clearly show the bimodality typical of massive galaxies, with blue and red peaks 
of $g-z \sim 0.9$ and $\sim 1.4$, respectively. This separation is twice as large as is typical 
of studies of GC systems in $V - I$ (e.g., Larsen \etal~2001; Kundu \& Whitmore 2001), due to 
the larger metallicity sensitivity of the $g-z$ baseline. However, a new result is that the red 
peak is clearly broader than the blue peak; at bright magnitudes ($z < 22$) there is little 
photometric error so this must be due to real color differences. To gauge the size of this 
effect, we fit a heteroscedastic normal mixture model to the M87 colors in the range $21 < z < 
22$. Subtracting a median photometric error of 0.02 mag in quadrature, we find $\sigma_{blue} 
\sim 0.07$ and $\sigma_{red} \sim 0.14$. These $\sigma$ values may be overestimates because of 
the presence of contaminants in the tails of the color distributions, but provide first-order 
estimates for investigation. Given the lack of evidence for significant age differences among 
bright GCs in massive early-type galaxies (Strader \etal~2005), it is reasonable to attribute 
the dispersion in $g-z$ entirely to metallicity.

To convert these dispersions into metallicity, we must find a relation between [M/H] and $g-z$. 
Jord{\' a}n \etal~(2004) used Bruzual \& Charlot (2003) models to find a linear relationship in the 
range $-2.3 \le$ [M/H] $\le +0.4$, however, the relation may be nonlinear for low metallicities. 
We fit a quadratic relation for [M/H] and $g-z$ using Maraston (2005) model predictions for 
$g-z$ for four metallicities ($-2.25, -1.35, -0.33, 0$) and Bruzual \& Charlot (2003) 
predictions at five metallicities ($-2.3, -1.7, -0.7, -0.4, 0$). Both sets of models assume a 13 
Gyr stellar population and a Salpeter initial mass function. The resulting fit is: [M/H] $= 
-8.088 + 9.081(g-z) - 2.524(g-z)^2$. Using this relation, the blue and red GC dispersions 
correspond to $1 \sigma$ metallicity ranges of ($-2.0$, $-1.4$) and ($-0.7$, $-0.1$), 
respectively. Thus, despite the wider color range of the red GCs, their logarithmic metallicity 
range appears no wider than that of the blue GCs due to the nonlinear relationship between $g-z$ 
and metallicity. A caveat is that this conclusion depends critically upon our assumed relation, 
which is likely to be most uncertain in the metal-poor regime where the stellar libraries of the 
models have few stars.

In M87, there is a clear enhancement of GCs at $z \sim 22.5$ with \emph{intermediate} colors, 
giving the CMD the appearance of a ``cosmic H''. This is illustrated more clearly in Figure 5,
which shows color histograms of GCs in the regions $22.2 < z < 22.7$ and $22.8 < z < 23.3$,
just below. Such a subpopulation of ``H" clusters is also present, albeit less 
clearly and slightly fainter, in NGC 4472. Since this is near the turnover of the GCLF (with the 
largest number of GCs per magnitude bin), it is difficult to acertain whether the enhancement is 
present at all colors or only in a narrow range. However, this subpopulation appears normal in all 
other respects. Defining a fiducial sample as lying in the range: $1.0 < g-z < 1.25$ and $22.2 < z 
< 22.7$, the sizes and radial distribution of these GCs lie between those of the blue and red GCs, 
though perhaps more similar to the blue ones. Visually they are indistinguishable from GCs of 
similar luminosity. With current data we cannot say how common these ``H'' GCs are in massive 
ellipticals, though their presence in NGC 4472 suggests that in the Virgo Cluster the phenomenon 
is not limited to M87. Here only 34 GCs fall into the limits defined above 
(though this is unlikely to define a complete sample); if this subpopulation scales with GC system 
richness, 10 or fewer might be expected in other luminous galaxies, rendering their detection 
difficult. At $z \sim 22.5$ spectroscopy of these GCs is feasible (though difficult), and could 
help establish whether their intermediate colors are due principally to metallicity or age, and 
whether they have kinematics distinct from the blue or red GCs. There may be a tail of these 
objects that extend to brighter magnitudes, but it is difficult to tell whether these are just 
outlying GCs in the normal blue or red subpopulations.

Also of interest is a group of anomalously bright GCs ($z \la 20$), which have a wide range in color ($0.9 
< g-z < 1.5$) and in some galaxies are separated from the bulk of the GC system by 0.1 mag or more. In 
M87, these GCs are on average $15 \%$ larger (with mean $r_h = 2.7$ pc) than GCs in the rest of the 
system, and have median galactocentric distances $\sim 10\%$ smaller ($\sim 5$ kpc) than the GC system as 
a whole. The dispersions in these properties appear consistent with the GC system as a whole, but with few 
bright clusters this is difficult to constrain. Some of these luminous GCs are likely in the tail of the 
normal blue and red subpopulations, but given the wide range in colors (including many with intermediate 
colors), small galactocentric radii, and the larger-than-average sizes, a portion may also be the stripped 
nuclei of dwarf galaxies---analogues of $\omega$ Cen in the Galaxy (Majewski \etal~2000). The color 
distribution of dE nuclei in this sample (see below) peaks at $g-z \sim 1.0-1.1$, consistent with the blue 
end of the intermediate-color objects. The surface brightness profiles of these objects resemble those of 
other GCs and do not have the exponential profiles seen in some ultra-compact dwarf galaxies (de Propris 
\etal~2005), though we note our size selection criterion for GCs would exclude some Virgo UCDs (Hasegan 
\etal~2005). Similar bright, intermediate-color GCs have also been found in the NGC 1399 (Dirsch 
\etal~2003), NGC 4636 (Dirsch, Schuberth, \& Richtler 2005), and NGC 1407 (Cenarro \etal~2005); they 
appear to be a common feature of gEs.

\subsubsection{The Blue Tilt}

A feature present in the CMDs of M87 and NGC 4649 is a \emph{tilt} of the color distribution of 
blue GCs, in the sense that the mean color of the blue GCs becomes redder with increasing 
luminosity. No such trend is apparent for the red GCs. A precise measure of this observation is 
challenging; due to the multiple subpopulations and ``H'' GCs, a direct linear fit is not viable. 
Instead, we divided the M87 GC candidates into four 0.5 mag bins in the range $20 < z < 22$ and 
one 0.4 mag bin ($22.8 < z < 23.2$, avoiding the ``H" GCs). To each of these bins we fit a 
heteroscedastic normal mixture model, and then fit a weighted linear model to the resulting blue 
peaks. This model is $g-z = -0.043 \, z + 1.848$; the slope is $4 \sigma$ significant. A fit to 
the corresponding red peaks is consistent with a slope of zero. These fits, as well as the binned 
values, are overplotted on the M87 CMD in Figure 2. Including a bin with the ``H'' GCs ($22 < z < 
22.8$) gives a slope which is (unsurprisingly) slightly more shallow ($-0.037$) but still 
significant. NGC 4649 has fewer GCs than M87 and appears to have no ``H'' GCs, so for this galaxy 
we fit three 1.0 mag bins in the range $20 < z < 23$. The resulting blue GC model is $g-z = -0.040 
\, z + 1.817$, which agrees very well with that of M87, and there is again no significant evidence 
for a nonzero red GC slope.

The smoothness of the change argues against stochastic stellar population changes (e.g., horizontal branch 
stars, blue stragglers) as the cause of the trend. If due to age, its size---$\sim 0.12-0.13$ mag in $g-z$ 
over $\sim 3$ mag in $z$---would require an unlikely age spread of $\sim 7-8$ Gyr at low metallicity using 
Maraston (2005) models. If due to metallicity, the color--metallicity relation derived above indicates the 
trend corresponds to a mean slope of $\sim 0.15-0.2$ dex/mag. For blue GCs in these galaxies, 
\emph{metallicity correlates with mass}. Interpretations of this surprising finding are discussed below; 
first we consider whether a bias in observation or analysis might be the cause.

Given that the correlation extends over a large range in GC luminosity, and is not seen for red 
GCs, selection bias (choosing redder GCs at bright magnitudes and bluer GCs at faint magnitudes) 
seems unlikely to be a factor. There is no significant correlation between GC luminosity and 
galactocentric radius, ruling out a radial variation in any quantity as a cause (e.g., dust). 
Together these facts also suggest that a systematic photometric error cannot be blamed.

To physically produce the observed correlation, either more massive GCs must have formed from more 
enriched gas, or individual GCs must have self-enriched. In the former picture, we could imagine 
blue GCs forming in proto-dwarf galaxies with varying metal enrichment. The essential problem is 
that there is no evidence that the GCLF varies strongly among dEs, as we would need the most 
metal-rich dEs to have few or no low-mass GCs to preserve the relation.

GC self-enrichment might explain the correlation, as more massive GCs could retain a larger 
fraction of supernovae (SNe) ejecta. The self-enrichment of GCs has been studied in some detail as 
a possible origin to the chemical inhomogeneities observed among stars in Galactic GCs (e.g., 
Smith 1987). Early works (e.g., Dopita \& Smith 1986) argued that only the most massive GCs could 
retain enough gas to self-enrich, but this depends critically on the assumed initial metal 
abundance of the proto-GC cloud and on the details of the cooling curve. Morgan \& Lake (1989) 
found that a more accurate cooling curve reduced the critical mass to $\ge 10^5 M_{\odot}$ in a 
``supershell" model, as suggested by Cayrel (1986). In the model of Parmentier \etal~(1999), 
proto-GC clouds are confined by a hot protogalactic medium, and this model in fact predicts an 
\emph{inverse} GC mass-metallicity relation, in which the most massive GCs are the most metal-poor 
(Parmentier \& Gilmore 2001). Clearly a wide range of models exist, and it is possible that with 
the appropriate initial conditions and physical mechanism a self-enrichment model of this sort can 
be made to work.

Another possibility is that the blue GCs formed inside individual dark matter (DM) halos. This 
scenario was first proposed by Peebles (1984), but fell into disfavor (Moore 1996) after studies 
of Galactic GCs found low mass-to-light ratios (Pryor \etal~1989) and tidal tails were observed 
around several GCs (e.g., Pal 5; Odenkirchen \etal~2003). Recently, Bromm \& Clarke (2002) and 
Mashchenko \& Sills (2005a,b) have used numerical simulations to argue that GCs with primordial DM 
halos could lose the bulk of the DM through either violent relaxation at early times or subsequent 
tidal stripping. If true, then a present-day lack of DM does not necessarily imply that GCs never 
had DM halos. It seems qualitatively plausible to produce the correlation in this context, but 
whether it could be sustained in detail requires additional simulation. Any such model need also 
be compared to the rather stringent set of other observations of blue GCs (some of which are not 
usually considered), including the lack of GC mass--radius and metallicity--galactocentric radius 
relations and the presence of a \emph{global} correlation between the mean metallicity of blue GCs 
and parent galaxy mass. In addition, since the Galaxy itself (and perhaps NGC 4472) show no 
obvious blue GC mass--metallicity relationship, variations among galaxies are needed.

It is also important to explain why the red GCs do not show such a relation. If the 
mass--metallicity relation was in terms of \emph{absolute} metallicity, then a small increase in 
metallicity ($0.01-0.02 Z_{\odot}$) could be visible among the blue GCs but not among the red GCs. 
Even if no weak relation exits, one cannot rule out a metallicity-dependent process that results 
in a relation only for the blue GCs even if both subpopulations formed the same way. Many other 
physical properties of the blue and red GCs are similar enough (e.g., GC mass functions, sizes) 
that it may be challenging to invoke completely different formation mechanisms. Some of the 
similarities could be due to post-formation dynamical destruction of low-mass or diffuse GCs, 
which might act to erase initial variations in some GC system properties. No consensus exists in 
the literature on the effectiveness of GC destruction in shaping the present-day GC mass function 
(e.g., Vesperini 2001; Fall \& Zhang 2001).

\subsection{Subpopulation Colors and Numbers}

The GC color distributions were modeled using the Bayesian program Nmix (Richardson \& Green 
1997), which fits normal heteroscedastic mixture models. The number of subpopulations is a free 
parameter (ranging to 10). For nearly all of the bright galaxies, two subpopulations were 
preferred; in no such galaxy was there a strong preference for a uni- or trimodal color 
distribution. Thus we report bimodal fits for these galaxies. While many of the dE color 
distributions visually appear bimodal, they generally had too few GCs to constrain the number of 
subpopulations with this algorithm. We adopted the following solution: we fit one peak to galaxies 
which only had GCs with $g-z < 1$; to the remaining galaxies we fit two peaks. A histogram of the 
dE GC colors (Figure 6) shows bimodality, which suggests that this approach is reasonable. Of 
course, some of the dEs have few GCs, so the peak locations may be quite uncertain.

Linear relationships have previously been reported between parent galaxy luminosity and the mean 
colors (peak/mode of a Gaussian fit) of the red (Larsen \etal~2001; Forbes \& Forte 2001) and blue 
(Strader, Forbes, \& Brodie 2004; Larsen \etal~2001; Lotz, Miller \& Ferguson 2004) 
subpopulations. Except for the massive Es, the individual subpopulations in this study have few 
GCs. Thus, errors on peak measurements are significant for most galaxies. Nevertheless, Figure 7 
shows that clear linear relationships are present for both the blue and red GCs over the full 
$\sim 6-7$ mag range in parent galaxy luminosity. These weighted relations are $g-z = -0.014 \, 
M_B + 0.642$ and $g-z = -0.053 \, M_B + 0.225$ for the blue and red GCs, respectively. The plotted 
error bars are standard errors of the mean. The fits exclude the gE NGC 4365, whose anomalous GC 
system has been discussed in detail elsewhere (Larsen, Brodie, \& Strader 2005, Brodie \etal~2005, 
Larsen \etal~2003, Puzia \etal~2002). For the red GCs there is a hint that the slope may flatten 
out for the faintest galaxies, but a runs test on the red residuals gave $p=0.33$ (and $p=0.45$ 
for the blue GCs), suggesting reasonable model fits. Many possible systematic errors could affect 
the faintest galaxies, e.g., the larger effects of contamination, and the uncertainty in the 
distances to individual galaxies, which could change their $M_B$ by $\sim 0.2-0.3$ mag. Thus, one 
cannot conclude that the GC color-galaxy luminosity relations are well-constrained at the faint 
end of our sample. However, they are consistent with extrapolations from brighter galaxies. Our 
results are also consistent with previous slope measurements: Larsen \etal~(2001) and Strader 
\etal~(2004) found that the $V-I$ red:blue ratio of slopes is $\sim 2$, while we find $\sim 3.7$ 
in $g-z$. This is consistent with $g-z \propto 2 \, (V-I)$, a rough initial estimate of color 
conversion (Brodie \etal~2005). There is at least one ongoing program to study Galactic GCs in the 
Sloan filter set which should improve this (and similar) conversions considerably.

It does appear that the residuals of the blue and red peak values are correlated; this is probably 
unavoidable when fitting heteroscedastic mixture models to populations which are not 
well-separated. Since the red and blue GC subpopulations clearly have different dispersions where 
this can be tested in detail, fitting homoscedastic models does not make sense. We experimented 
with fitting two-component models with the variances fixed to the mean value for the brightest 
galaxies, for which the large number of GCs (at least partially) breaks the degeneracy between 
peak location and dispersion. The slopes of the resulting blue and red relations are similar to 
those found using the above approach: $-0.012$ and $-0.057$, respectively. However, the blue peak 
values for many of the galaxies appear to be artificially high---this may be because galaxies less 
massive than gEs have smaller intra-subpopulation metallicity spreads. Thus we have chosen to 
leave the original fits as our final values.

The very existence of red GCs in faint galaxies with $M_B \sim -15$ to $-16$ is an interesting and 
somewhat unexpected result. In massive early-type galaxies and many spiral bulges, the number of 
red GCs normalized to spheroid luminosity is approximately constant (Forbes, Brodie, \& Larsen 
2001). This suggests that red GCs formed along with the spheroidal field stars at $\sim$ constant 
efficiency. However, many properties of dEs (e.g., surface brightness profiles, M/L ratios, 
spatial/velocity distribution, stellar populations) suggest that their formation mechanism was 
different from massive Es (e.g., Kormendy 1985, though see Graham \& Guzm{\' a}n 2003 for a 
different view). A continuity of red GC properties between Es and at least some of the dEs in our 
sample could imply either that their formation mechanisms were more similar than expected or that 
red GCs are formed by a self-regulating, local process that can occur in a variety of contexts.

The mixture modeling also returns the number of GCs in each subpopulation. In Figure 8 we plot the 
fraction of blue GCs vs.~galaxy luminosity. There is a general trend (with a large scatter) for an 
increasing proportion of blue GCs with decreasing galaxy luminosity. The average fraction of blue 
GCs for the dEs is $\sim 0.7$, with many galaxies having no red GCs. The fraction asymptotes to 
$\sim 0.4 - 0.5$ for the gEs, but since these data cover only the central, more red-dominated part 
of their GC systems, the global fraction is likely higher (e.g., in a wide-field study of three gEs 
by Rhode \& Zepf 2004, the blue GC fraction ranged from 0.6--0.7.). The GC systems of the dEs fall 
entirely within the ACS field of view (see discussion below), so their measured blue GC fractions 
are global.

These results show clearly that the classic correlation between GC metallicity/color and galaxy luminosity 
(Brodie \& Huchra 1991) is a combination of two effects: the decreasing ratio of blue to red GCs with 
increasing galaxy luminosity, and, more importantly, the GC color-galaxy luminosity relations which exist 
for \emph{both} subpopulations. There does not seem to be a single ``primordial'' GC color--galaxy 
luminosity relation, as assumed in the accretion scenario for GC bimodality (e.g., C{\^ o}t{\' e}, Marzke, 
\& West 2002)

\section{Luminosity Functions and Nuclei}

Many previous works have found (e.g., Harris 1991, Secker 1992, Kundu \& Whitmore 2001, Larsen 
\etal~2001) that the GCLF in massive galaxies is well-fit by a Gaussian or $t_5$ distribution with 
similar properties among well-studied galaxies: $M_V \sim -7.4$, $\sigma_V \sim 1.3$.  However, 
the shape among dwarf galaxies is poorly known. Individual galaxies have too few GCs for a robust 
fit, and thus a composite GCLF of many dwarfs is necessary. Using ground-based imaging, Durrell 
\etal~(1996) found that the turnover of the summed GCLF of 11 dEs in Virgo was $\sim 0.4 \pm 0.3$ 
mag fainter the M87 turnover. Lotz \etal~(2001) presented HST/WFPC2 snapshot imaging of 51 dEs in 
Virgo and Fornax; the summed GCLF in Lotz \etal~is not discussed, but a conference proceeding using the same data 
(Miller 2002) suggests $M_V \sim -7.4$ and $-7.3$ for Virgo and Fornax, respectively.

We study the dE GCLF through comparison to gEs (VCC 1316--M87, VCC 1226--NGC 4472, and VCC 
1978--NGC 4649) previously found to have ``normal" GCLFs (Larsen \etal~2001). For dwarfs we 
constructed a summed GCLF of all 37 dEs ($M_B < -18.2$ in our sample). Since the individual 
distance moduli of the dEs are unknown, this GCLF could have additional scatter because of the 
range of galaxy distances; we find below that this appears to be quite a small effect. The 
individual GC systems of dEs are quite concentrated: most GCs are within $30-40\arcsec$, beyond 
which the background contamination rises sharply (size measurements are unreliable at these faint 
magnitudes). Thus for the dEs we selected only those GCs within $30\arcsec$ of the center of the 
galaxy.

In the previous sections we used photometric and structural cuts to reduce the number of background galaxies and 
foreground stars interloping in our GC samples. However, these same cuts cannot be directly used to fit GCLFs, 
since these measurements become increasingly inaccurate for faint GCs (which may then be incorrectly removed). 
For example, applying only the color cut ($0.5 < g-z < 2.0$) vs.~the full selection criteria to M87 changes the 
number of GCs with $z < 23.5$ by 16\% (1444 vs.~1210) and shifts the peak of the GCLF by $\sim 0.4$ mag, which is 
quite significant.

We directly fit $t_5$ distributions to the three gE GCLFs using using the code of Secker \& Harris (1993), 
which uses maximum-likelihood fitting and incorporates photometric errors and incompleteness (Gaussian fits 
gave similar results within the errors; note that $\sigma_{gauss} = 1.29 \, \sigma_{t5}$). These GCLFs only 
have a color cut applied: $0.5 < g-z < 2.0$ for M87 and NGC 4472 and $0.85 < g-z < 2.0$ for NGC 4649 (since 
star-forming regions in the nearby spiral NGC 4647 strongly contaminate the blue part of the CMD). The 
results are given in Table 2. We fit both total populations and blue and red GCs separately, using the color 
cuts given in the table. Similar to what is commonly seen in $V$-band GCLFs (e.g., Larsen \etal~2001), the 
$g$ turnovers are $\sim 0.3-0.4$ mag brighter for the blue GCs than the red GCs.  This is predicted for 
equal-mass/age turnovers separated by $\sim 1$ dex in metallicity (Ashman, Conti \& Zepf 1995). However, in 
$z$ the blue GCs are only slightly brighter than the red GCs (the mean difference is negligible, but the blue 
GCs are brighter in M87 and NGC 4472 and fainter in NGC 4649, which may be biased slightly faint because of 
contamination). This difference between $g$ and $z$ is qualitatively consistent with stellar population 
models: Maraston (2005) models predict that equal-mass 13 Gyr GCs with [M/H] = $-1.35$ and $-0.33$ will have 
$\Delta g = 0.6$ and $\Delta z = 0.2$; the $z$ difference is one-third of the $g$ difference. This effect is 
probably due to the greater sensitivity of $g$ to the turnoff region and the larger number of metal lines in 
the blue. The errors for the blue GCLF parameters may be slightly larger than formally stated because of the 
presence of the ``H" GCs discussed previously. Variations in this feature and in the blue tilt among galaxies 
could represent a fundamental limitation to the accuracy to using \emph{only} the blue GCLF turnover as a 
standard candle (see, e.g., Kissler-Patig 2000). However, the total peak locations themselves are quite 
constant, with a range of only 0.03 mag in $g$ and 0.05 mag in $z$. At least for the well-populated old GC 
systems of gEs, the GCLF turnover appears to be an accurate distance indicator whose primary limitation is 
accurate photometry.

Unfortunately, we cannot directly fit the composite dE GCLF as for the gEs---it is far too contaminated. 
Instead, we must first correct for background objects. As noted above, the GC systems of the dEs are very 
centrally concentrated. Thus we can use the outer regions of the dE images as a fiducial background. We 
defined a background sample in the radial range $1.1-1.25 \arcmin$ (corresponding to 5.4--6.2 kpc at a 
distance of 17 Mpc), then subtracted the resulting $z$ GCLF in 0.1 mag bins from the central $30 \arcsec$ 
sample with the appropriate areal correction factor. We confined the fit to candidates with $0.7 < g-z < 
1.25$ because of the small color range of the dEs. The resulting fit (performed as above) gave a turnover of 
$M_z = -8.14\pm0.14$, compared to the weighted mean of $M_z = -8.19$ for the gEs. These are consistent. Since 
the $z$ GCLFs are being used, the overall color differences between the GC systems should have little effect 
on the peak locations. The dispersion of the dEs ($\sigma_z = 0.74$) is less than that of the gEs ($\sigma_z 
= 1.03$); this is probably partially due to the smaller color range of the dEs. It also indicates that the 
range of galaxy distances is probably not significant, consistent with the projected appearance of many of 
the dEs near the Virgo cluster core (Binggeli \etal~1987).

That the peak of the GCLF appears to be the same for both gEs and dEs is perhaps puzzling. Even if both 
galaxy types had similar primordial GCLFs, analytic calculations and numerical simulations of GC evolution in 
the tidal field of the dEs suggest that dynamical friction should destroy high-mass GCs within several Gyr 
(Oh \& Lin 2000, Lotz \etal~2001). Tidal shocks (which tend to destroy low-mass GCs) are more 
significant in massive galaxies, so it is unlikely that the loss of high-mass GCs in dEs could be balanced by 
a corresponding amount of low-mass GC destruction to preserve a constant turnover. \emph{ab initio} 
variations in the GCLF between dEs and gEs are another possibility, but cannot be constrained at present.

If we see no evidence in the dE GCLF turnover for dynamical friction, does this provide significant evidence 
against this popular scenario for forming dE nuclei? In order study the relationship between the nuclei and GCs 
of our sample galaxies, we performed photometry and size measurements on all of the dE nuclei. The procedures 
were identical to those described above for the GCs, except that aperture corrections were derived through 
analytic integration of the derived King profile, since many of the nuclei were large compared to typical GC 
sizes. The properties of the nuclei are listed in Table 3.

In Figure 9 we plot parent galaxy $M_B$ vs.~$M_z$ of the nucleus. Symbol size is proportional to 
nuclear size. The overplotted dashed line represents Monte Carlo simulations of nuclear formation 
through dynamical friction of GCs from Lotz \etal (2001). These assume 5 Gyr of orbital decay 
(implicitly assuming the GCs are $\sim 5$ Gyr old); we have converted their $M_V$ to $M_z$ using 
stellar populations models of Maraston (2005). Lotz \etal~found that the expected dynamical 
friction was inconsistent with observations, since the observed nuclei were fainter than predicted 
by their simulations. They argued that some additional process was needed to oppose dynamical 
friction (e.g., tidal torques).

At first glance, this is supported by Figure 9---for galaxies in the luminosity range $-15 < M_B < -17$, 
there is a $\sim 3$ mag range in nuclear luminosity. But most of the bright nuclei are quite small, while 
those much fainter than predicted from the simulations are generally larger (by factors of 2--4). In 
addition, the small nuclei tend to be redder than the large ones; some of the large faint nuclei are blue 
enough that they require some recent star formation. While not a one-to-one relation, it seems plausible that 
these luminosity and size distributions reflect two different channels of nuclear formation: small bright red 
nuclei are formed by dynamical friction, while large faint blue nuclei are formed by a dissipative process 
with little or no contribution from GCs. The red nuclei have typical colors of $g-z \sim 1.1$, consistent 
with those of the red GCs in dEs. This implies that red GCs (rather than blue GCs) would need to be the 
primary ``fuel" for nuclei built by dynamical friction, at least among the brigher dEs. The formation of blue 
nuclei could happen as the dE progenitor entered the cluster, or during tidal interactions that drive gas to 
the center of the galaxy (i.e., harassment, Mayer \etal~2001). Present simulations do not resolve the central 
parts of the galaxy with adequate resolution to estimate the size of a central high-density component, but 
presumably this will be possible in the future.

It is unclear how to square the hypothesis that some nuclei are built by dynamical friction with the GCLF 
results from above. Perhaps the small $\sigma_z$ of the composite dE (compared to the gEs) is additional 
evidence of GC destruction. Our background correction for the dE GCLF is potential source of systematic error 
in the turnover determination. Deeper ACS photometry for several of the brighter Virgo dEs could be quite 
helpful, since it could give both better rejection of background sources and more accurate SBF distances to 
the individual dEs.

\section{Specific Frequencies}

For the Es in our sample, the ACS FOV covers only a fraction of their GC system, and no robust 
conclusions about their total number of GCs can be drawn without uncertain extrapolation. The GC 
systems of fainter Es and dEs are less extended and fall mostly or entirely within ACS pointings. 
As previously discussed, $S_N$ is often used as a measure of the richness of a GC system. In fact, 
Harris \& van den Bergh (1981) originally defined $S_N$ in terms of the \emph{bright} GCs in a 
galaxy: the total number was taken as the number brighter than the turnover doubled. The 
justification for this procedure was that the faint end of the GCLF was often ill-defined, with 
falling completeness and rising contamination. In addition, for a typical log-normal GCLF centered 
at $M_V = -7.4 \sim 3 \times 10^5 M_{\odot}$, 90\% or more of the total GC system mass is in the 
bright half of the LF.

Many of the dEs in this study have tens or fewer detected GCs, making accurate measurements of the shape 
of the LF impossible. Thus counting only those GCs brighter than the turnover is unwise; small variations 
in the peak of the GCLF among galaxies (either due to intrinsic differences or the unknown galaxy 
distances) could cause a large fraction of GC candidates to be included or excluded. As discussed in \S 2, 
we chose $z = 23.5$ as the magnitude limit for our study. This is $\sim 0.5$ mag beyond the turnover of 
the composite dE GCLF discussed in the previous section ($M_z \sim -8.1$, $\sigma_z \sim 0.7$) at a 
nominal Virgo Cluster distance of 17 Mpc. For such galaxies our magnitude cutoff corresponds to $\sim 75$\% 
completeness. Thus, to get total GC populations for the dEs, we divided the number of GCs brighter than $z 
= 23.5$ by this factor (for the few dEs with individual distance moduli, we integrated the LF to find the 
appropriate correction factor). Finally, as discussed in \S 4, the photometric and structural cuts used to 
reduce contaminants also remove real GCs. Using M87 as a standard, 16.2\% of real GCs with $z < 23.5$ were 
falsely removed; we add back this statistical correction to produce our final estimate of the total GC 
population. Due to the small radial extent of the dE GC systems (primarily within a projected galactocentric 
radius of $\sim 30-40\arcsec$) no correction for spatial coverage is applied.

We calculated $B$-band $S_N$ using the absolute magnitudes in Table 1. Unless otherwise noted, all 
$S_N$ refer to $B$-band values. These $S_{N}$ can be converted to the standard $V$-band $S_N$ by 
dividing our values by a factor $10^{0.4(B-V)}$; a typical dE has $B-V = 0.8$, which corresponds 
to a conversation factor $\sim 2.1$. $S_{N}$ is plotted against $M_B$ in Figure 10. In this figure 
the dE,N galaxies are filled circles and the dE,noN galaxies are empty circles. The symbol size is 
proportional to the fraction of blue GCs.

Miller \etal~(1998) found that dE,noNs had, on average, lower $S_N$ than dE,Ns, and that for both 
classes of dE there was an inverse correlation between $S_N$ and galaxy luminosity. We do not see 
any substantial difference between dE,noN and dE,N galaxies. There does appear to be a weak 
correlation between $S_N$ and $M_B$ in our data (with quite large $S_N$ for some of the faintest 
galaxies), but there is a large spread in $S_N$ for most luminosities. The $S_N$ does not appear 
to depend strongly on the fraction of blue GCs, so differing mixes of subpopulations cannot be 
responsible for the $S_N$ variations. Thus the observed values correspond to total blue GC 
subpopulations which vary by a factor of up to $\sim 10$ at a given luminosity. There appear
to be two sets of galaxies: one group with $S_{N} \sim 2$, the other group ranging from $S_{N}
\sim 5-20$ and a slight enhancement at $S_{N} \sim 10$. We only have 
one galaxy in common with Miller \etal~(VCC 9), but our $S_N$ for this single galaxy is consistent 
with theirs within the errors. We note again that background contamination remains as a systematic 
uncertainty for total GC populations.

It is tempting to argue that these groups represent different formation channels for dEs, e.g., 
the fading/quenching of dIrrs, harassment of low-mass spirals, or simply a continuation of the E 
sequence to fainter magnitudes. In some scenarios for blue GC formation (e.g., Strader \etal~2005; 
Rhode \etal~2005), galaxies with roughly similar masses in a given environment would be expected 
to have similar numbers of blue GCs per unit mass of their DM halos. Variations in $S_{N}$ at 
fixed luminosity would then be due to differences in the efficiency of converting baryons to 
stars---either early in the galaxy's history (due to feedback) or later, due to galaxy 
transformation processes as described above. Two pieces of observational evidence should help to 
differentiate among the possibilities. First, GC kinematics can be used directly to constrain dE 
mass-to-light ratios (e.g., Beasley \etal~2005). It will be interesting to see whether GC 
kinematics are connected to the apparent dichotomy of rotating vs.~non-rotating dEs (Pedraz 
\etal~2002, Geha \etal~2002, 2003, van Zee \etal~2004). Second, more detailed stellar population 
studies of dEs are needed, especially any method which (unlike integrated light spectroscopy) can 
break the burst strength--age degeneracy. A rather ad hoc alternative is that the efficiency of 
blue GC formation varies substantially among Virgo dEs, but this cannot be constrained at present. 
It is interesting that the fraction of blue GCs appears unrelated both to the nucleation of the dE 
and the $S_N$ variations; this suggests that whatever process leads to red GC formation in dEs is 
independent of these other factors.

\section{Summary}

We have presented a detailed analysis of the GC color and luminosity distributions of several gEs and of the 
colors, specific frequencies, luminosity functions, and nuclei of a large sample of dEs. The most interesting 
feature in the gEs M87 and NGC 4649 is a correlation between mass and metallicity for individual blue GCs. 
Self-enrichment is a plausible interpretation of this observation, and could suggest that these GCs once 
possessed dark matter halos (which may have been subsequently stripped). Among the other new features 
observed are very luminous ($z \ga 20$) GCs with intermediate to red colors. These objects are slightly 
larger than typical GCs and may be remnants of stripped dwarf galaxies. Next, we see an intermediate-color 
group of GCs which lies near the GCLF turnover and in the gap between the blue and red GCs. Also, the color 
spread among the red GCs is nearly twice that of the blue GCs, but because the relation between $g-z$ and 
metallicity appears to be nonlinear, the $1 \sigma$ dispersion in metallicity ($\sim 0.6$ dex) may be the 
same for both subpopulations.

The peak of the GCLF is the same in the gEs and a composite dE, modulo uncertainties in background 
subtraction for the dEs. This observation may be difficult to square with theoretical expectations that 
dynamical friction should deplete massive GCs in less than a Hubble time, and with the properties of dE 
nuclei. There appear to be two classes of nuclei: small bright red nuclei consistent with formation by 
dynamical friction of GCs, and larger faint blue nuclei which appear to have formed by a dissipative process 
with little contribution from GCs.

Though dominated by blue GCs, many dEs appear to have bimodal color distributions, with significant red GC 
subpopulations. The colors of these GCs form a continuity with those of more massive galaxies; both the mean 
blue and red GC colors of dEs appear consistent with extrapolations of the GC color--galaxy luminosity 
relations for luminous ellipticals. We confirm these relations for both blue and red GC subpopulations. While 
previous works found an inverse correlation between dE $S_N$ and galaxy luminosity, we find little support 
for such a relation. There is a large scatter in $S_N$ among dEs and some evidence for two separate groups of 
galaxies: dEs with $S_N \sim 2$, and dEs with large GC systems that have $S_N$ ranging from $\sim 5-20$ with 
median $S_N \sim 10$. However, these $S_N$ variations do not appear to be connected to the presence of a 
nucleus or the fraction of red GCs. Our findings suggest multiple formation channels for Virgo dEs.

\acknowledgements

We thank an anonymous referee for comments that considerably improved the manuscript.
We acknowledge support by the National Science Foundation through Grant AST-0206139 and a Graduate 
Research Fellowship (J.~S.). This research has made use of the NASA/IPAC Extragalactic Database 
(NED), which is operated by the Jet Propulsion Laboratory, California Institute of Technology, 
under contract with the National Aeronautics and Space Administration. We thank Graeme Smith for 
useful conversations.

\newpage

\begin{figure}
\plotone{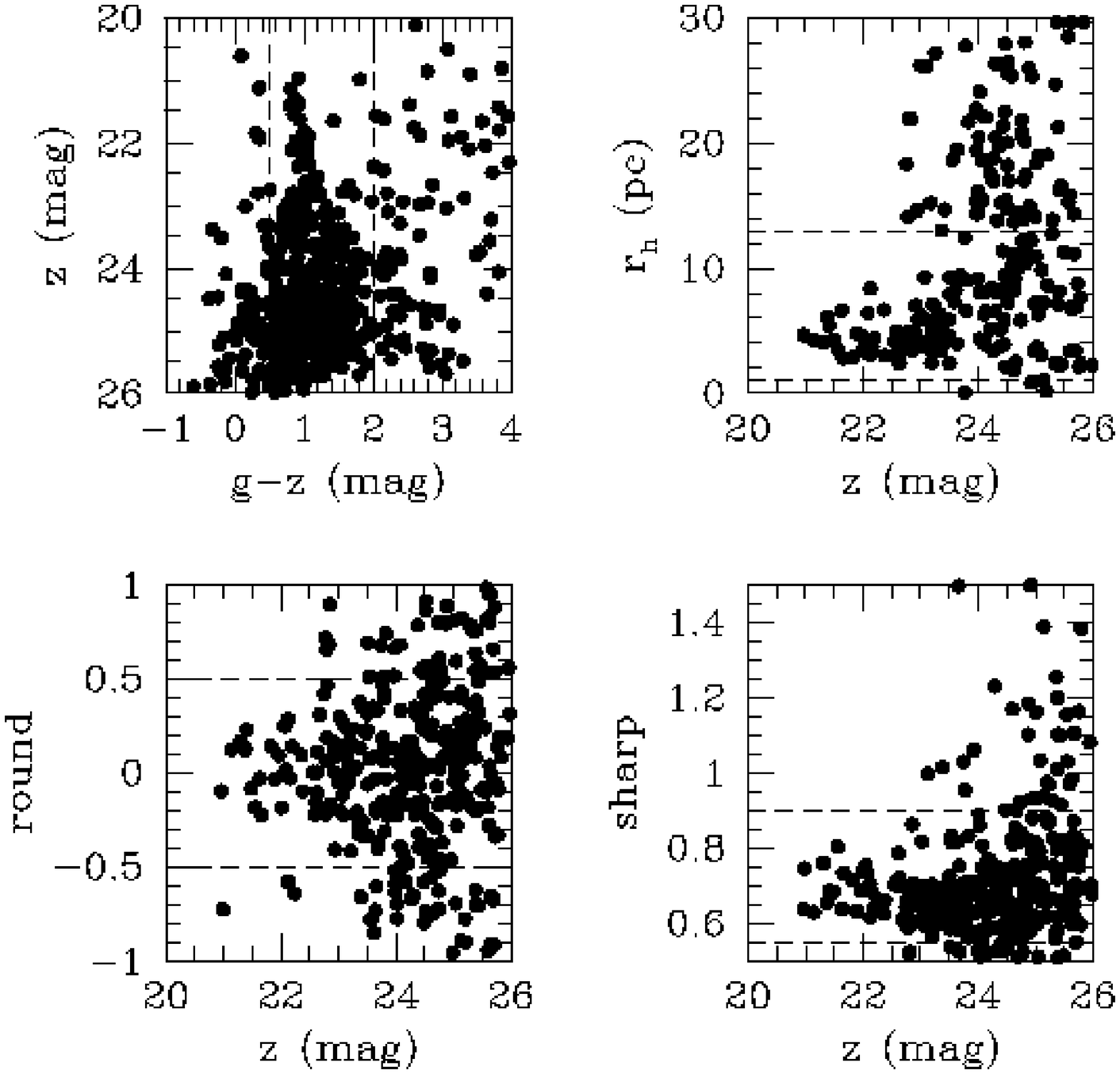}
\figcaption[f1.eps]{\label{fig:fig1} Plots of GC candidates in VCC 1087 that indicate our
structural cuts. Clockwise from top left, these are: $z$ vs.~$g-z$ color-magnitude diagram,
half-light radius ($r_h$) vs.~$z$, sharp vs.~$z$, and round vs.~$z$. All plots except the
color-magnitude diagram have a color selection of $0.5 < g-z < 2.0$.}
\end{figure}

\begin{figure}
\plotone{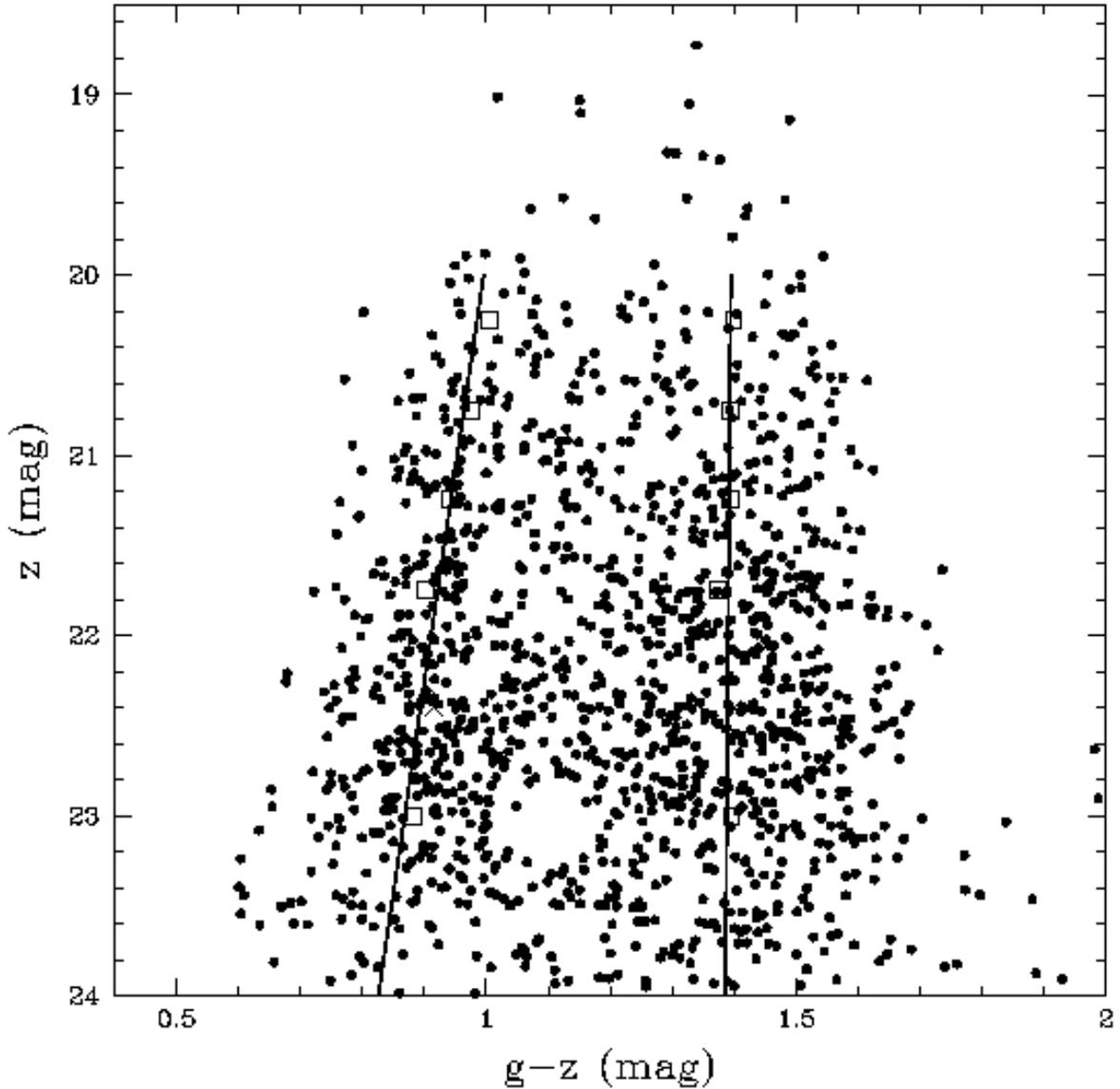}
\figcaption[f2.eps]{\label{fig:fig2} 
The $z$ vs.~$g-z$ color-magnitude diagram for GC candidates in M87 (VCC 1316). The overplotted
boxes are peak locations for individual bins, and the lines are fits to these binned values. The two
crosses show peak values for the bin with ``H'' GCs, and are not included in the plotted fits.
The ``tilt" of the blue GCs suggests a correlation between metallicity and mass for these clusters. See
text for more details.}
\end{figure}

\newpage

\begin{figure}
\plotone{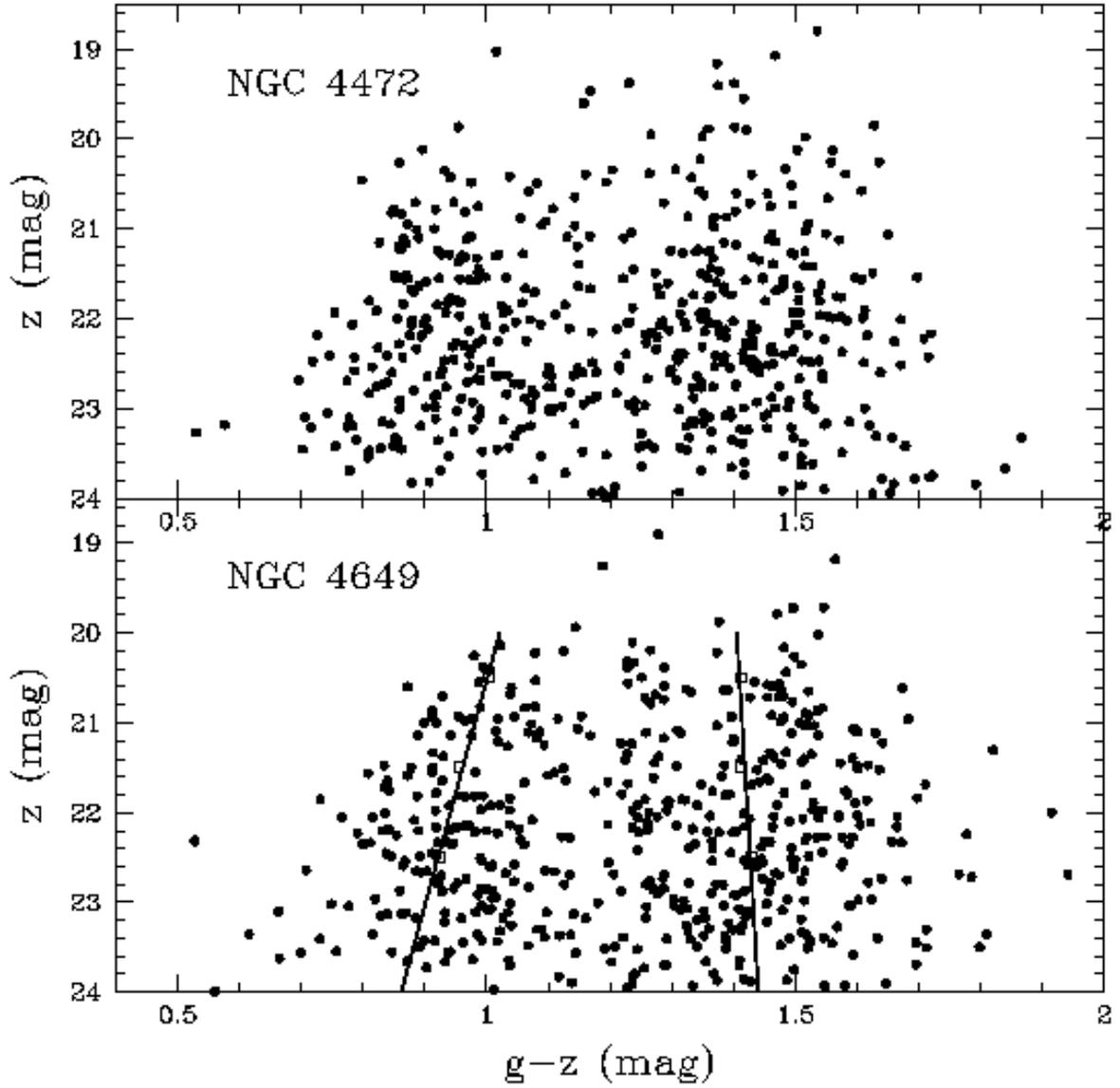}
\figcaption[f3.eps]{\label{fig:fig3} The $z$ vs.~$g-z$ color-magnitude diagrams for GC
candidates in NGC 4472 (VCC 1226) and NGC 4649 (VCC 1978). Symbols for NGC 4649 are as in
Figure 1.}
\end{figure}

\newpage

\begin{figure}
\plotone{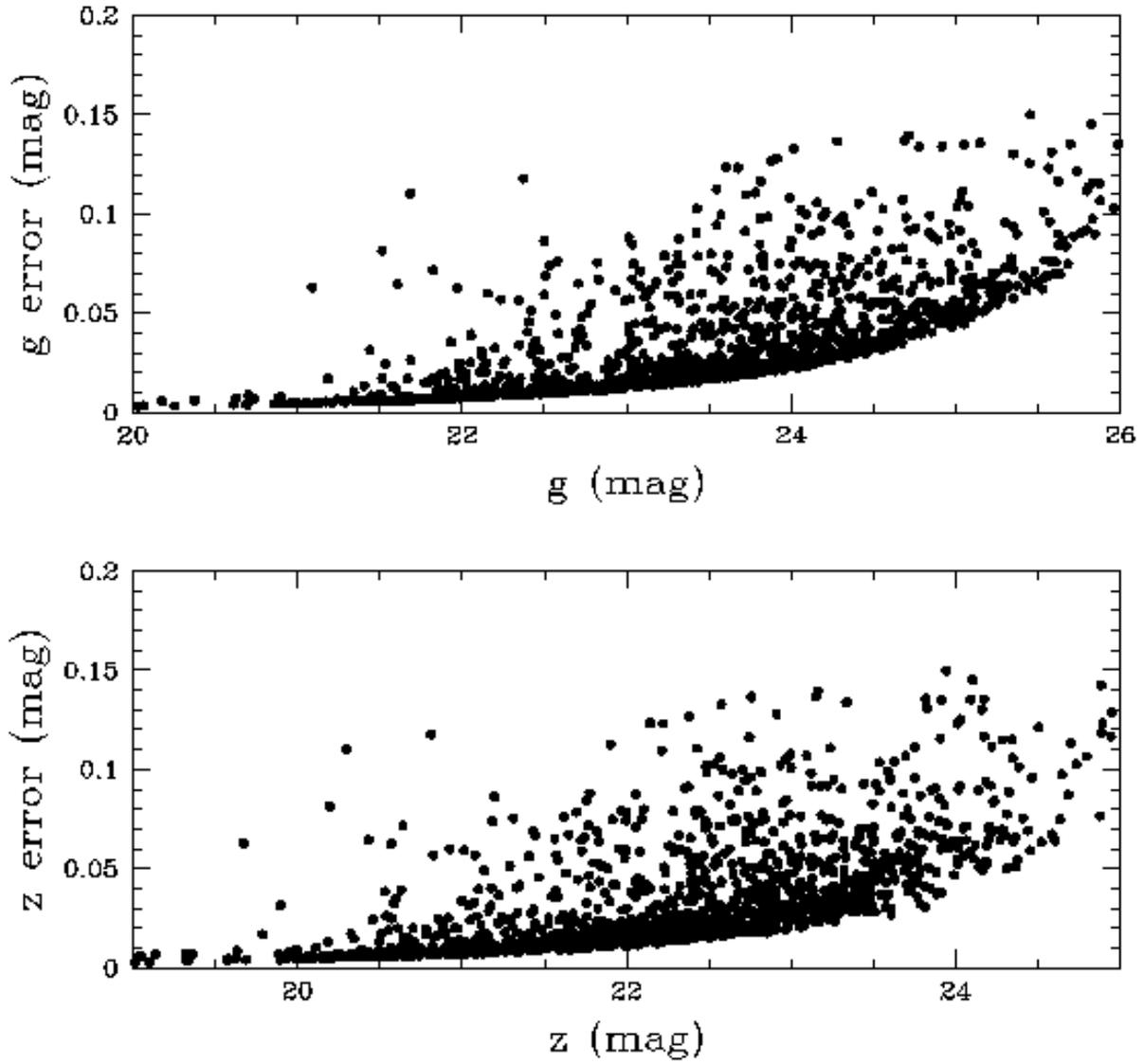}
\figcaption[f4.eps]{\label{fig:fig4} $g$ and $z$ photometric error vs.~magnitude for GC
candidates in M87.}
\end{figure}

\newpage

\begin{figure}
\plotone{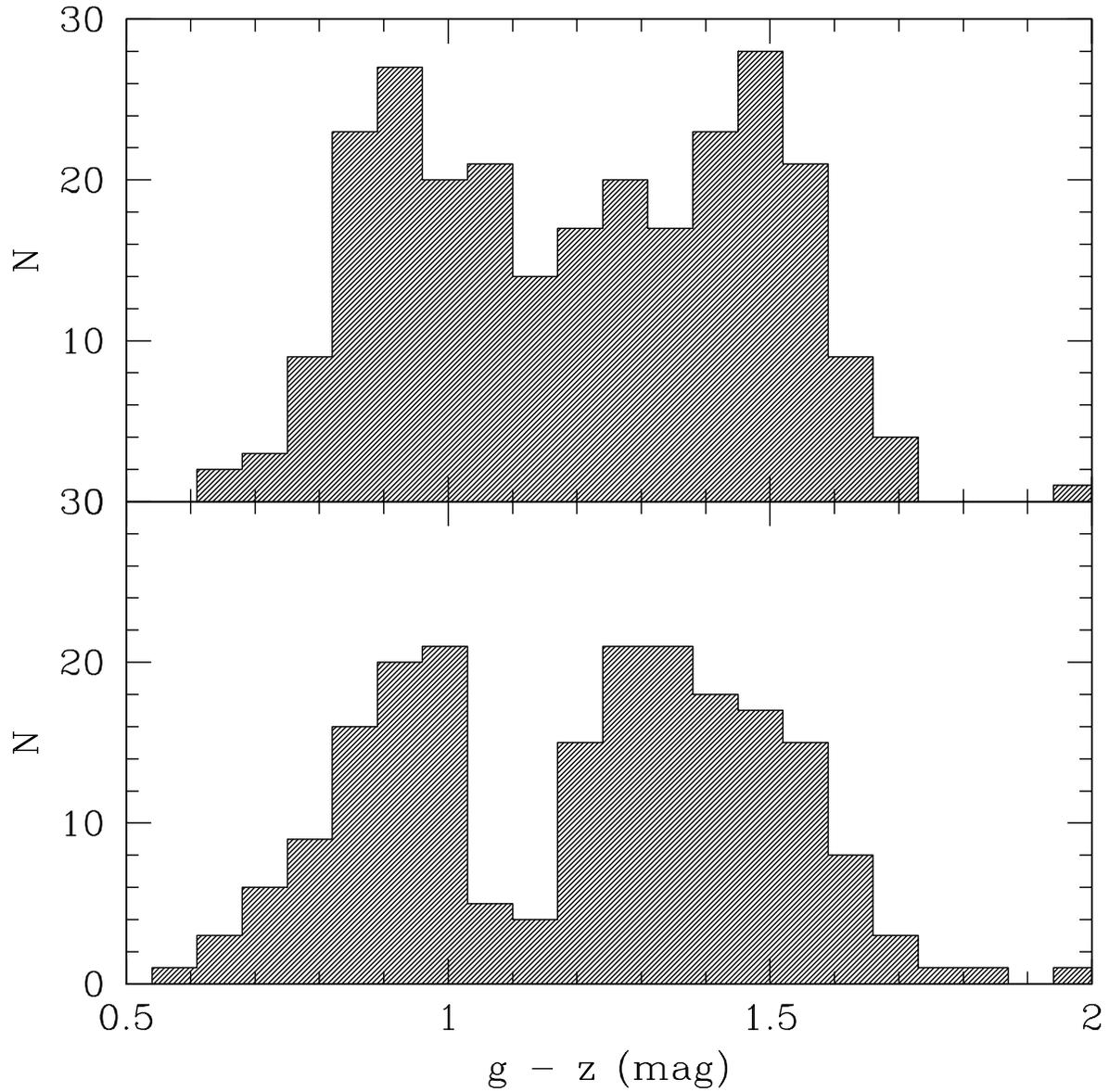}
\figcaption[f5.eps]{\label{fig:fig5} $g-z$ histograms for GCs in the ranges $22.2 < z < 22.7$ (``H" GCs) and 
$22.8 < z < 23.3$, just below. The lower panel shows the normal gap between the blue and red GC subpopulations, while the upper 
panel shows how the ``H" GCs have filled the gap in.}
\end{figure}

\newpage

\begin{figure}
\plotone{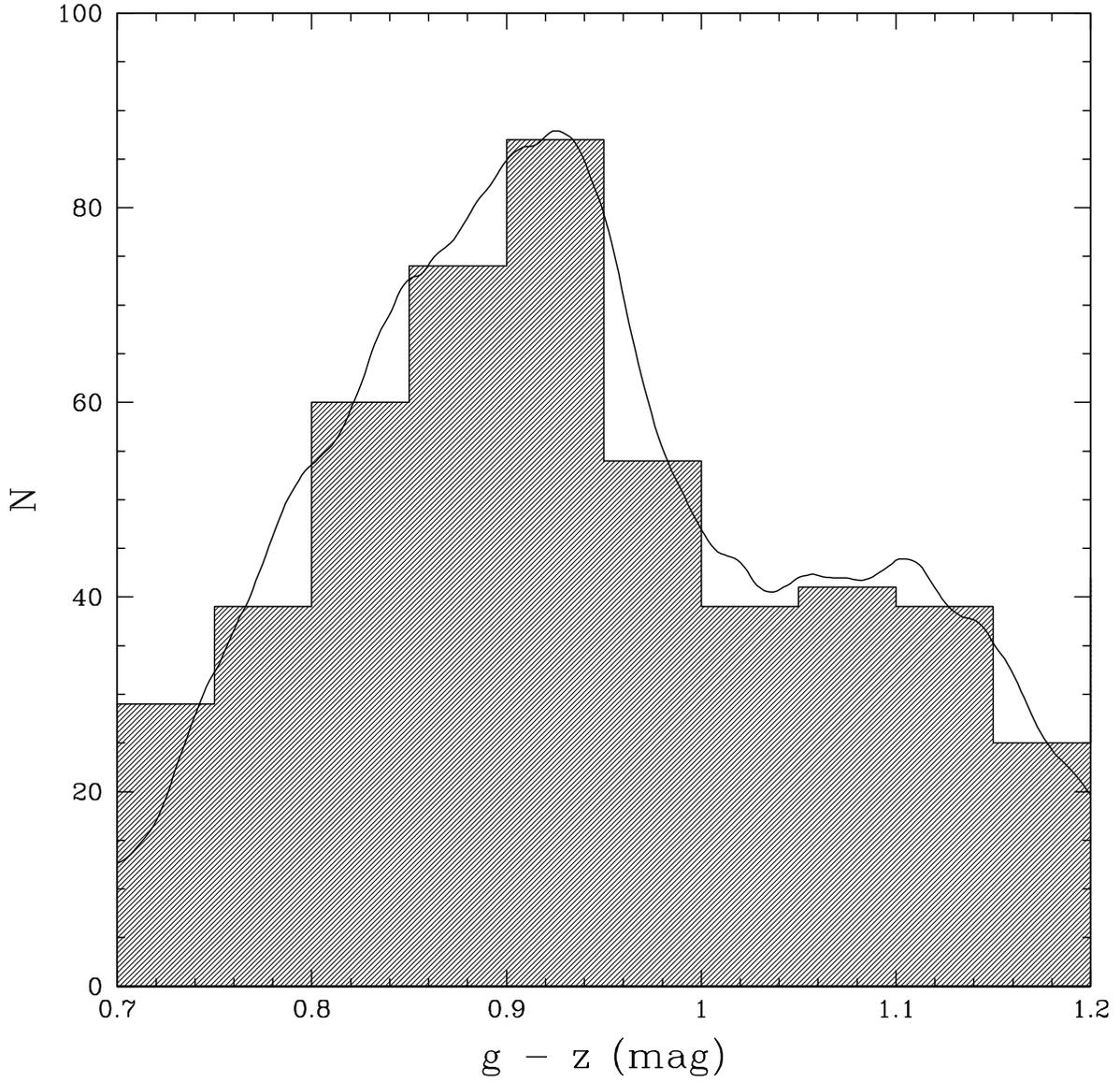}
\figcaption[f6.eps]{\label{fig:fig6} $g-z$ histogram of dE GC candidates. A density estimate using
an Epanechnikov kernel and a bin width of 0.02 mag is overplotted, showing evidence for bimodality.}
\end{figure}

\newpage

\begin{figure}
\plotone{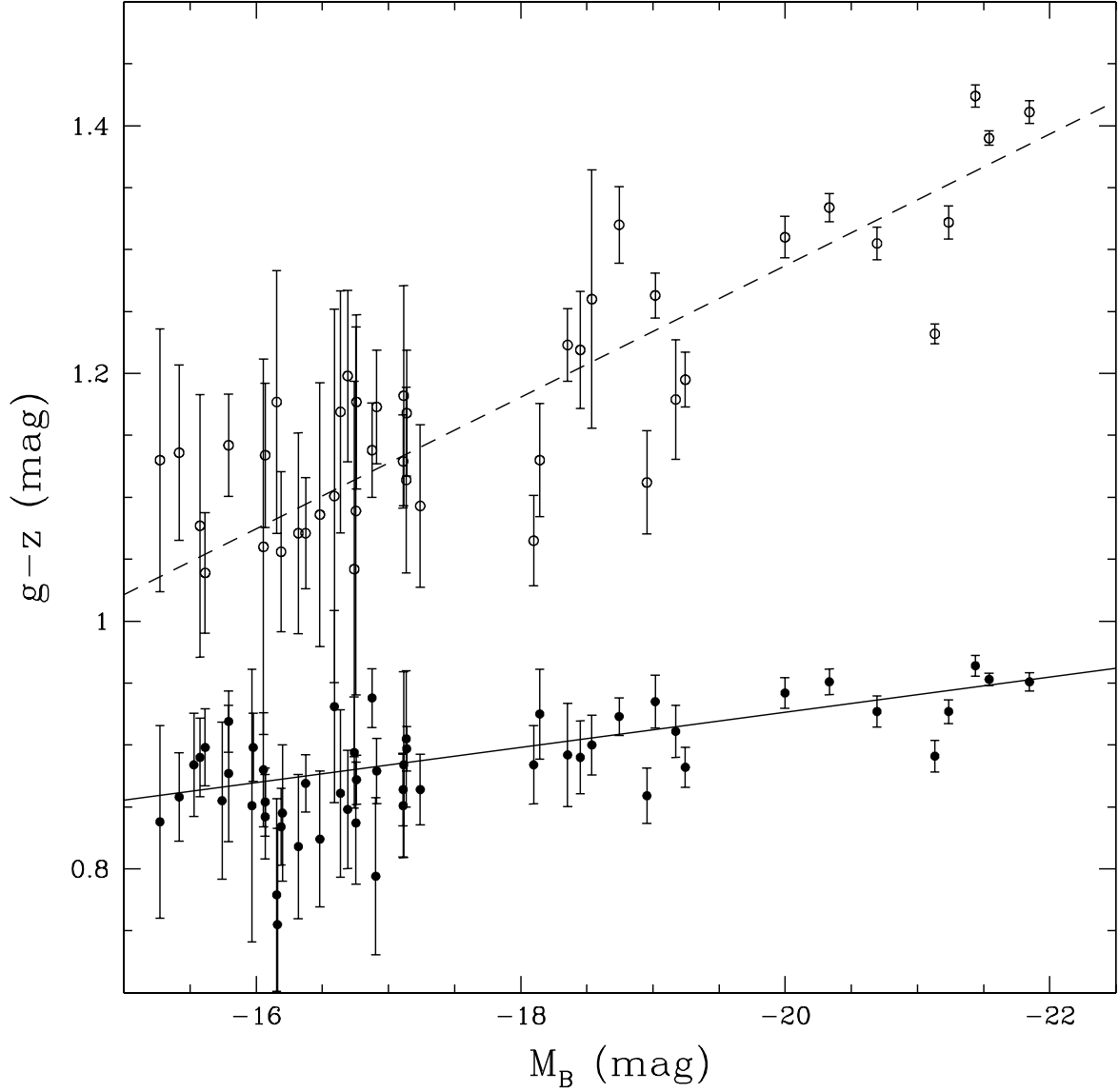}
\figcaption[f7.eps]{\label{fig:fig7} Mean $g-z$ color of blue GCs (filled circles) and red GCs (open circles) 
vs.~parent galaxy $M_B$. The lines represent weighted linear fits.}
\end{figure}

\newpage

\begin{figure}
\plotone{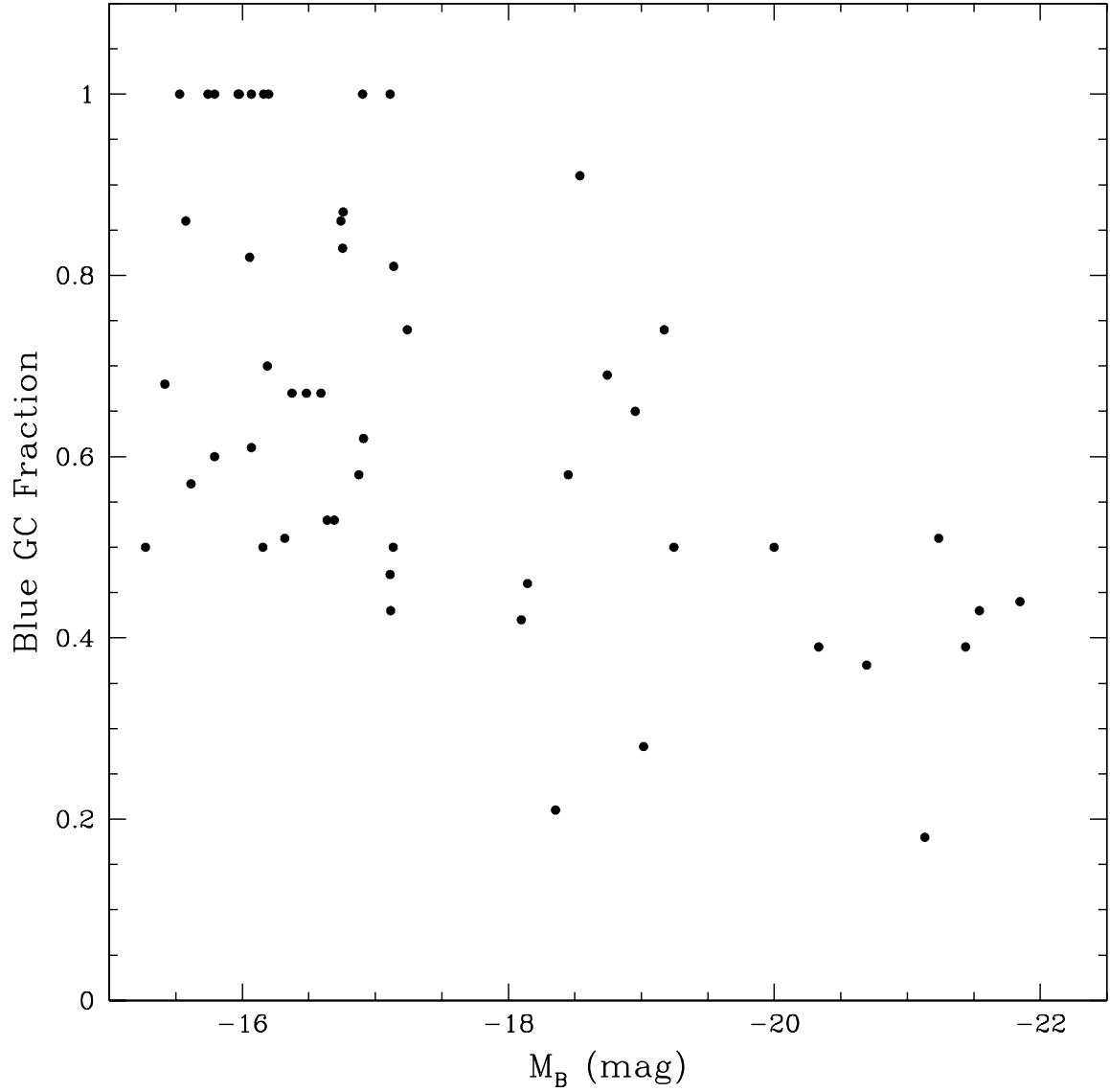}
\figcaption[f8.eps]{\label{fig:fig8} The fraction of blue GCs vs.~parent galaxy $M_B$. As a general trend, 
luminous galaxies have more red GCs, though the \emph{overall} fraction is poorly constrained for massive 
galaxies in our sample due to small areal coverage.}
\end{figure}

\newpage

\begin{figure}
\plotone{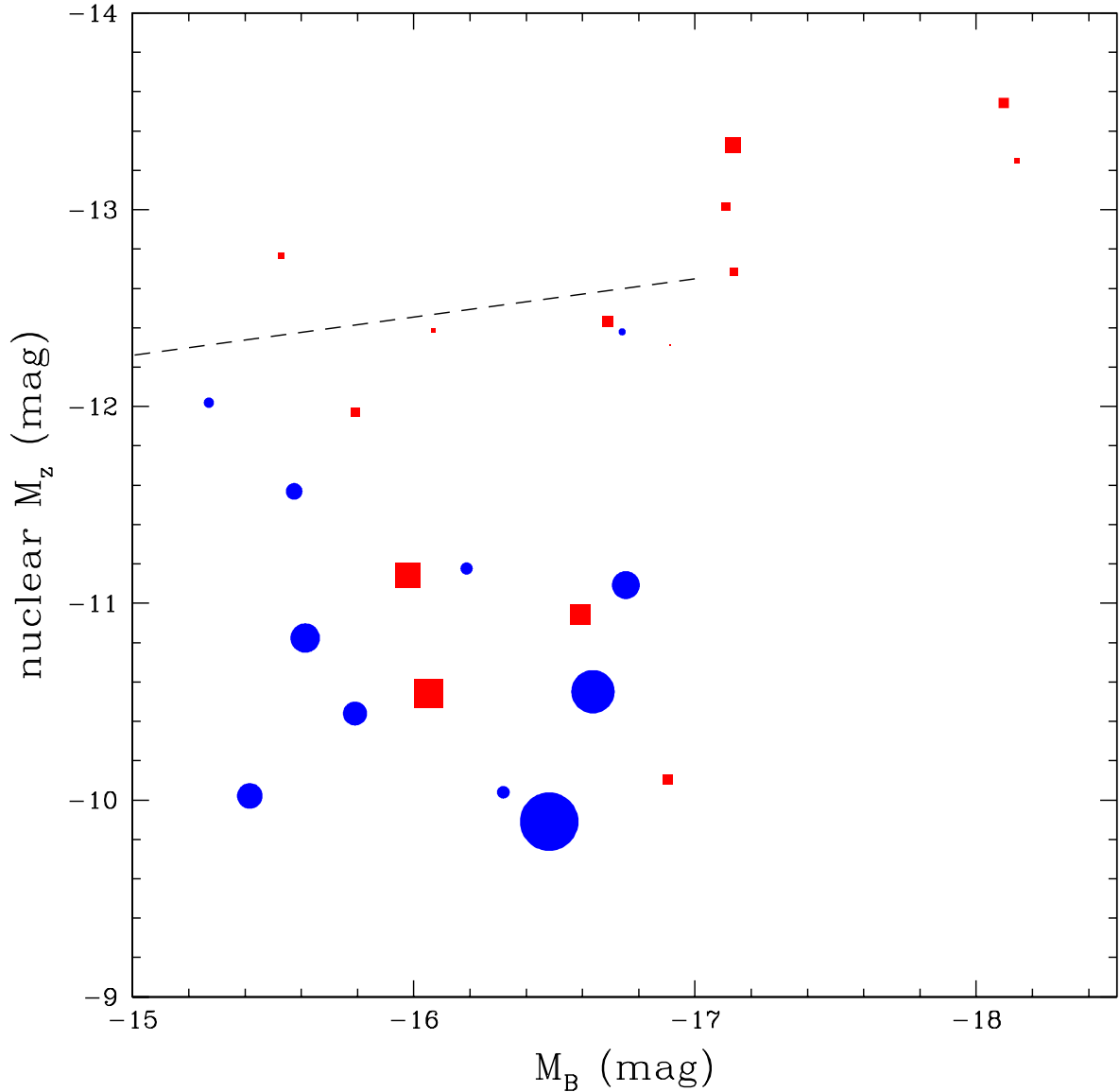}
\figcaption[f9_color.eps]{\label{fig:fig9} $M_z$ of dE nuclei vs.~parent galaxy $M_B$. The size of the points 
is proportional to the size of the nucleus. Blue circles are nuclei with $g-z < 1$; red squares are nuclei with $g-z \ge 
1$. The overplotted line is a simulated trend of the creation of dE nuclei through dynamical friction of GCs (Lotz 
\etal~2001). There appear to be two classes of dE nuclei: small bright red nuclei which are approximately consistent 
with  the numerical predictions of dynamical friction, and large faint blue nuclei which form (at least 
partially) through a dissipative process.}
\end{figure}

\newpage

\begin{figure}
\plotone{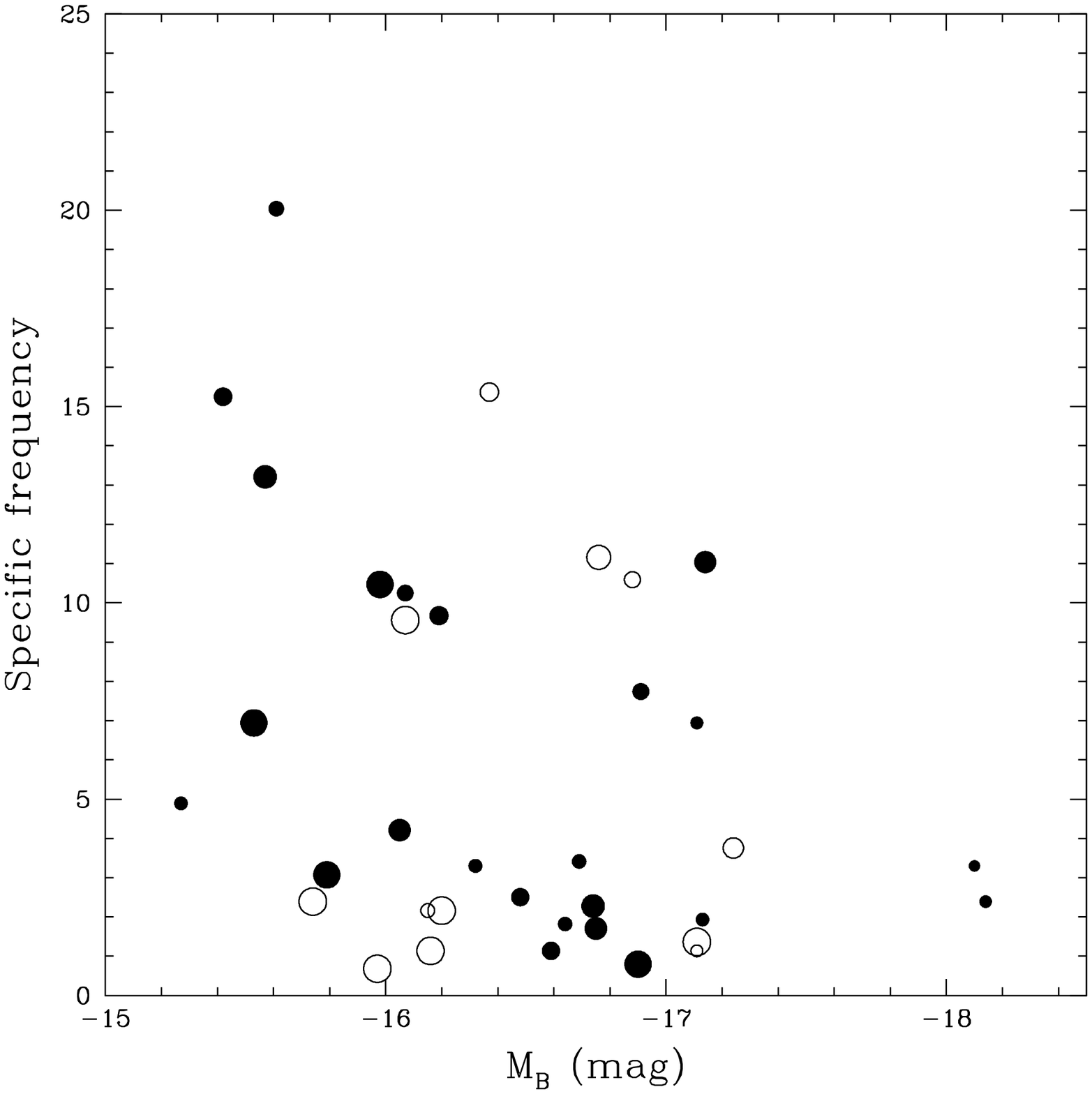}
\figcaption[f10.eps]{\label{fig:fig10} Specific frequency ($S_N$) of dE,Ns (filled cirles) and dE,noNs 
(open circles) vs.~parent galaxy $M_B$. The size of the points is proportional to the fraction of blue GCs. 
There is only a weak trend of increasing $S_N$ with decreasing $M_B$, and no substantial difference between 
the two galaxy classes.}
\end{figure}

\newpage

\begin{deluxetable}{lrrrrrrrrrrr}
\tablewidth{0pt} 
\tabletypesize{\scriptsize}
\tablecaption{Globular Cluster System Data
        \label{tab:gcdata}}
\tablehead{Galaxy  & Alt. Name & dE,N? & $M_B$ & $m-M$\tablenotemark{a} & Blue peak & Red peak & No. GCs    & Blue frac. & Red frac. & 
$S_N$ \\ 
           VCC     &   NGC/IC  &       & (mag) & (mag) & (mag)     & (mag)    & ($z < 23.5$)  &         &                 }
\startdata

1226 & N4472 & \nodata & $-$21.85 & 31.06 & 0.951 & 1.411 & 482 & 0.44 & 0.56 & \nodata  \\
1316 & N4486 &\nodata  & $-$21.54 & 31.03 & 0.953 & 1.390 & 1180 & 0.43 & 0.57 & \nodata  \\
1978 & N4649 & \nodata & $-$21.44 & 31.13 & 0.964 & 1.424 & 451 & 0.39 & 0.61 & \nodata  \\
763 & N4374 & \nodata &  $-$21.24 & 31.32 & 0.927 & 1.322 & 259 & 0.51 & 0.49 & \nodata  \\
731 & N4365 & \nodata &  $-$21.13 & 31.55 & 0.891 & 1.232 & 419 & 0.18 & 0.82 & \nodata \\
1903 & N4621 & \nodata & $-$20.69 & 31.31 & 0.927 & 1.305 & 209 & 0.37 & 0.63 & \nodata  \\
1632 & N4552 & \nodata & $-$20.33 & 30.93 & 0.951 & 1.334 & 288 & 0.39 & 0.61 & \nodata  \\
1231 & N4473 & \nodata & $-$20.00 & 30.98 & 0.942 & 1.310 & 160 & 0.50 & 0.50 & \nodata \\
1279 & N4478 & \nodata & $-$19.24 & 31.29 & 0.882 & 1.195 & 92 & 0.50 & 0.50 & \nodata  \\
1025 & N4434 & \nodata & $-$19.17 & 32.14 & 0.911 & 1.179 & 37 & 0.74 & 0.26 & \nodata  \\
1664 & N4564 & \nodata & $-$19.02 & 30.88 & 0.935 & 1.263 & 95 & 0.28 & 0.72 & \nodata  \\
828 & N4387 & \nodata &  $-$18.95 & 31.65 & 0.859 & 1.112 & 37 & 0.65 & 0.35 & \nodata  \\
2000 & N4660 & \nodata & $-$18.74 & 30.54 & 0.923 & 1.320 & 76 & 0.69 & 0.31 & \nodata  \\
1321 & N4489 & \nodata & $-$18.54 & 31.26 & 0.900 & 1.260 & 23 & 0.91 & 0.09 & \nodata  \\
1630 & N4551 & \nodata & $-$18.45 & 31.19 & 0.890 & 1.219 & 24 & 0.58 & 0.42 & \nodata  \\
1146 & N4458 & \nodata & $-$18.35 & 31.19 & 0.892 & 1.223 & 33 & 0.21 & 0.79 & \nodata  \\
1422 & I3468 & Y &       $-$18.14 & 31.64 & 0.925 & 1.130 & 20 & 0.46 & 0.54 & 2.4  \\
1261 & N4482 & Y & 	 $-$18.10 & 31.54 & 0.884 & 1.065 & 29 & 0.42 & 0.58 & 3.3  \\
9 & I3019 & N & 	 $-$17.24 & 31.00 & 0.864 & 1.093 & 20 & 0.74 & 0.26 & 3.8  \\
1087 & I3381 & Y & 	 $-$17.14 & 31.33 & 0.897 & 1.168 & 46 & 0.81 & 0.19 & 11.0  \\
856 & I3328 & Y & 	 $-$17.13 & 31.28 & 0.905 & 1.114 & 8 & 0.50 & 0.50 & 1.9  \\
575 & N4318 & N & 	$-$17.11 & \nodata & 0.884 & 1.182 & 5 & 0.43 & 0.57 & 1.4  \\
1871 & I3653 & N & 	$-$17.11 & 30.84 & 0.851 & \nodata & 7 & 1.00 & 0.00 & 1.2  \\
1910 & I809 & Y & 	$-$17.11 & \nodata & 0.864 & 1.129 & 30 & 0.47 & 0.53 & 6.9  \\
1861 & I3652 & Y & 	$-$16.91 & \nodata & 0.879 & 1.173 & 28 & 0.62 & 0.38 & 7.7  \\
543 & \nodata & Y & 	$-$16.90 & \nodata & 0.794 & \nodata & 3 & 1.00 & 0.00 & 0.8  \\
1431 & I3470 & N & 	$-$16.88 & \nodata & 0.938 & 1.138 & 37 & 0.58 & 0.42 & 10.6  \\
1528 & I3501 & N & 	$-$16.76 & \nodata & 0.872 & 1.177 & 35 & 0.87 & 0.13 & 11.2  \\
1355 & I3442 & Y & 	$-$16.75 & 30.92 & 0.837 & 1.089 & 6 & 0.83 & 0.17 & 1.7  \\
437 & \nodata & Y & 	$-$16.74 & \nodata & 0.894 & 1.042 & 7 & 0.86 & 0.14 & 2.3  \\
2019 & I3735 & Y & 	$-$16.69 & \nodata & 0.848 & 1.198 & 10 & 0.53 & 0.47 & 3.4  \\
33 & I3032 & Y & 	$-$16.64 & \nodata & 0.861 & 1.169 & 5 & 0.53 & 0.47 & 1.8  \\
200 & \nodata & Y & 	$-$16.59 & \nodata & 0.931 & 1.101 & 3 & 0.67 & 0.33 & 1.1  \\
1488 & I3487 & Y & 	$-$16.48 & \nodata & 0.824 & 1.086 & 6 & 0.67 & 0.33 & 2.5  \\
1545 & I3509 & N & 	$-$16.37 & \nodata & 0.869 & 1.071 & 34 & 0.67 & 0.33 & 15.4 \\ 
1895 & \nodata & Y & 	$-$16.32 & \nodata & 0.818 & 1.071 & 7 & 0.51 & 0.49 & 3.3  \\
1857 & I3647 & N & 	$-$16.20 & \nodata & 0.845 & \nodata & 4 & 1.00 & 0.00 & 2.2 \\
1075 & I3383 & Y & 	$-$16.19 & \nodata & 0.834 & 1.056 & 18 & 0.70 & 0.30 & 9.7 \\
1627 & \nodata & N & 	$-$16.16 & \nodata & 0.755 & \nodata & 2 & 1.00 & 0.00 & 1.1 \\
1948 & I798 & N & 	$-$16.15 & \nodata & 0.779 & 1.177 & 4 & 0.50 & 0.50 & 2.2 \\
230 & I3101 & Y & 	$-$16.07 & \nodata & 0.842 & 1.134 & 17 & 0.61 & 0.39 & 10.3 \\
1440 & \nodata & N & 	$-$16.07 & \nodata & 0.854 & \nodata & 16 & 1.00 & 0.00 & 9.6 \\
2050 & I3779 & Y & 	$-$16.05 & \nodata & 0.880 & 1.060 & 7 & 0.82 & 0.14 & 4.2 \\
1828 & I3635 & Y & 	$-$15.98 & \nodata & 0.898 & \nodata & 16 & 1.00 & 0.00 & 10.5 \\
1993 & \nodata & N & 	$-$15.97 & \nodata & 0.851 & \nodata & 1 & 1.00 & 0.00 & 0.7  \\
1407 & I3461 & Y & 	$-$15.79 & \nodata & 0.919 & 1.142 & 33 & 0.60 & 0.40 & 25.5  \\
1886 & \nodata & Y & 	$-$15.79 & \nodata & 0.877 & \nodata & 4 & 1.00 & 0.00 & 3.1 \\
1743 & I3602 & N & 	$-$15.74 & \nodata & 0.855 & \nodata & 3 & 1.00 & 0.00 & 2.4 \\
1539 & \nodata & Y & 	$-$15.61 & \nodata & 0.898 & 1.039 & 22 & 0.57 & 0.43 & 20.0 \\
1185 & \nodata & Y & 	$-$15.57 & \nodata & 0.890 & 1.077 & 14 & 0.86 & 0.14 & 13.2  \\
1826 & I3633 & Y & 	$-$15.53 & \nodata & 0.884 & \nodata & 7 & 1.00 & 0.00 & 6.9  \\
1489 & I3490 & Y & 	$-$15.42 & \nodata & 0.858 & 1.136 & 14 & 0.68 & 0.32 & 15.3 \\
1661 & \nodata & Y & 	$-$15.27 & & 0.838 & 1.130 & 4 & 0.50 & 0.50 & 4.9  \\

\enddata

\tablenotetext{a}{Distance modulus, if available from Tonry \etal~(2001) or Jerjen \etal~(2004). Otherwise the Tonry \etal~
mean value of 17 Mpc ($m-M$ = 31.15) is used.}

\end{deluxetable}

\begin{deluxetable}{lrrrrrrr}
\tablewidth{0pt}  
\rotate
\tablecaption{Luminosity Function Fits\tablenotemark{a}
        \label{tab:lf}}
\tablehead{Galaxy & Band & Total $\mu$ & Total $\sigma$ & Blue $\mu$ & Blue $\sigma$ & Red $\mu$ & Red $\sigma$\\
                 &      & (mag) & (mag) & (mag) & (mag) & (mag) & (mag)}

\startdata

M87             & $g$ & $-7.08\pm0.04$ & $0.73\pm0.02$ & $-7.36\pm0.04$ & $0.65\pm0.03$ & $-6.93\pm0.04$ & $0.82\pm0.04$  \\
\nodata         & $z$ & $-8.19\pm0.06$ & $1.07\pm0.07$ & $-8.23\pm0.08$ & $0.92\pm0.08$ & $-8.16\pm0.09$ & $1.16\pm0.10$  \\
NGC 4472        & $g$ & $-7.05\pm0.04$ & $0.85\pm0.04$ & $-7.31\pm0.08$ & $0.82\pm0.06$ & $-6.86\pm0.06$ & $0.90\pm0.05$  \\
\nodata         & $z$ & $-8.21\pm0.06$ & $0.97\pm0.05$ & $-8.29\pm0.09$ & $0.92\pm0.07$ & $-8.15\pm0.09$ & $1.02\pm0.07$  \\
NGC 4649        & $g$ & $-7.08\pm0.04$ & $0.86\pm0.03$ & $-7.24\pm0.08$ & $0.86\pm0.06$ & $-6.99\pm0.06$ & $0.87\pm0.04$  \\
\nodata         & $z$ & $-8.16\pm0.08$ & $1.06\pm0.06$ & $-8.05\pm0.15$ & $1.05\pm0.10$ & $-8.22\pm0.09$ & $1.05\pm0.07$  \\
\hline
gE mean         & $g$ & $-7.07$         & 0.81         & $-7.30$       & 0.78           & $-6.93$       & 0.86 \\
\nodata         & $z$ & $-8.19$         & 1.03         & $-8.19$       & 0.96           & $-8.18$       & 1.08 \\

\enddata

\tablenotetext{a}{The quantities in this table are calculated using the distance moduli in Table 1. The 
errors reflect only the statistical errors in the maximum-likelihood 
fitting process. The color cuts used are---M87: $g-z=1.10$, NGC 4472/NGC 4649: $g-z=1.15$. $\mu$ and 
$\sigma$ are the mean and dispersion of the fitted t5 functions.}

\end{deluxetable}

\begin{deluxetable}{lrrrr}
\tablewidth{0pt}
\tablecaption{Properties of Dwarf Elliptical Nuclei\tablenotemark{a}
        \label{tab:denuc}}
\tablehead{Galaxy & $M_g$\tablenotemark{b} & $M_z$\tablenotemark{b} & $g-z$ & $r_h$ \\
            VCC   & (mag) & (mag) & (mag) & (pc)}

\startdata

1261 & $-$12.49 & $-$13.54 & 1.05 & 4.9 \\
856 & $-$12.24 & $-$13.33 & 1.07 & 7.1 \\
1422 & $-$11.92 & $-$13.25 & 1.20 & 2.2 \\
1910 & $-$11.74 & $-$13.02 & 1.14 & 3.9 \\
437 & $-$11.52 & $-$12.38 & 0.87 & 2.4 \\
1087 & $-$11.41 & $-$12.69 & 1.27 & 3.6 \\
1826 & $-$11.41 & $-$12.77 & 1.13 & 2.9 \\
1861 & $-$11.39 & $-$12.31 & 1.04 & 0.6 \\
230 & $-$11.20 & $-$12.39 & 1.03 & 2.1 \\
2019 & $-$11.15 & $-$12.43 & 1.10 & 5.2 \\
1661 & $-$11.01 & $-$12.02 & 0.91 & 3.4 \\
1407 & $-$10.93 & $-$11.97 & 1.04 & 4.1 \\
1185 & $-$10.60 & $-$11.57 & 0.96 & 5.4 \\
1075 & $-$10.35 & $-$11.18 & 0.81 & 4.0 \\
1355 & $-$10.11 & $-$11.09 & 0.98 & 9.1 \\
1539 & $-$10.06 & $-$10.82 & 0.80 & 9.6 \\
1828 & $-$10.04 & $-$11.14 & 1.01 & 11.5 \\
200 & $-$9.86 & $-$10.94 & 1.18 & 9.5 \\
33 & $-$9.61 & $-$10.55 & 0.88 & 14.2 \\
2050 & $-$9.45 & $-$10.54 & 1.05 & 13.3 \\
1886 & $-$9.35 & $-$10.44 & 0.91 & 7.9 \\
1489 & $-$9.06 & $-$10.02 & 0.82 & 8.4 \\
1488 & $-$8.99 & $-$9.89 & 0.75 & 19.2 \\
1895 & $-$8.99 & $-$10.04 & 0.95 & 4.1 \\
543 & $-$8.91 & $-$10.10 & 1.13 & 4.8 \\

\enddata

\tablenotetext{a}{The quantities in this table are calculated using the distance moduli in Table 1.}
\tablenotetext{b}{The photometric errors in these quantities are $\le 0.01$ mag for most galaxies and $\le 0.02$ mag
for the remainder.}

\end{deluxetable}

\end{document}